\documentclass{svproc}
\usepackage{amsmath,graphicx}

\usepackage{url}

\begin{document}
\mainmatter              
\title{Exploring Transfer Learning for Low Resource Emotional TTS}
%
%
\author{No\'e Tits, Kevin El Haddad and Thierry Dutoit}

\authorrunning{No\'e Tits et al.} 
%
%
\institute{University of Mons, \\
  Numediart Institute, \\
	Mons, Belgium 7000 \\
\email{\{noe.tits, kevin.elhaddad, thierry.dutoit\}@umons.ac.be}
}

\maketitle              

\begin{abstract}
During the last few years, spoken language technologies have known a big improvement thanks to Deep Learning. However Deep Learning-based algorithms require amounts of data that are often difficult and costly to gather. Particularly, modeling the variability in speech of different speakers, different styles or different emotions with few data remains challenging. In this paper, we investigate how to leverage fine-tuning on a pre-trained Deep Learning-based TTS model to synthesize speech with a small dataset of another speaker. Then we investigate the possibility to adapt this model to have emotional TTS by fine-tuning the neutral TTS model with a small emotional dataset.


\end{abstract}

\section{INTRODUCTION}
The current state of the art of Text-to-Speech (TTS) synthesis is based on deep learning algorithms. These systems are now capable of producing natural human-like speech.

There are more and more deep learning-based TTS systems developed. The Merlin toolkit~\cite{merlin-16-wu} has played an important part in this development, first with simple architectures based on stacks of Fully Connected layers and then with more complex ones such as Recurrent Neural Networks (RNNs). 

Most recent TTS systems, such as Wavenet~\cite{wavenet-16-vandenoord}, Tacotron~\cite{tacotron-17-wang}, WaveRNN~\cite{wavernn-18-Kalchbrenner}, Char2Wav~\cite{char2wav-17-sotelo} and Deep Voice~\cite{deep-voice-17-arik},  achieve excellent results in terms of naturalness. However they require tens of hours of speech data and a lot of computational power. A first system that aims to synthesize speech with few computational power is Deep Convolutional TTS~\cite{dctts-17-tachibana} (DCTTS). This system is only based on Convolutional Neural Networks (CNNs) and avoids using RNNs known to be difficult to train due to the vanishing gradient issue during gradient descent~\cite{vanishing-gradient-98-Hochreitera}. In their experiments, the authors of DCTTS were able to train their model in 15 hours using a standard PC with two GPUs, resulting in nearly acceptable speech synthesis.

Moreover it is difficult to have a fine control on speech quality and emotional content with such systems, while this has become an important challenge in speech synthesis. Here again, data availability is an issue. Indeed, high quality speech datasets with emotional content needed for speech synthesis are quite difficult to collect. The amount of data available is therefore relatively limited compared to what deep learning algorithms require to converge. 

Promising methods to tackle the problem of quantity of data are those related to knowledge transfer such as transfer learning~\cite{transfer_learning-2010-pan}, fine-tuning and multi-task learning. These techniques have prooved useful in various applications of deep learning.


In the field of Motion Capture and Analysis,~\cite{mocap-img-17-laraba} mapped a motion sequence to an RGB image to be able to use a CNN pre-trained for image classification in their motion classification task. They showed that fine-tuning the CNN on their motion data improved classification results.

In a previous work of ours~\cite{asr-based-features-18-tits}, we used an neural Automatic Speech Recognition (ASR) as a feature extractor for emotion recognition. We showed that the mapping between speech and text learned by the ASR system contains information useful for emotion recognition.

In the TTS field as well, transfer learning is being investigated. In~\cite{tacotron3-18-ye}, they successfully transfered knowledge from a model trained to discriminate between speakers to a multi-speaker TTS model.

These examples motivates our interest to investigate the use of knowledge transferability between models. 
Our goal is to use this to tackle the inherent problem of the amount of data needed in deep learning in the case of emotional TTS which is a topic of growing interest. We can cite~\cite{tacotron2-17-shen} that experimented an unsupervised learning technique to change prosody of synthesized sentences with style tokens or~\cite{emo_tacotron-17-lee} that modified Tacotron's architecture to synthesize speech given emotional labels.
In this paper we explain how to leverage fine-tuning on deep learning-based TTS systems to synthesize emotional speech with a small emotional speech dataset.

\section{SYSTEM}

\begin{figure*}[h]
\centering
\includegraphics[scale=0.2, width = 1\textwidth]{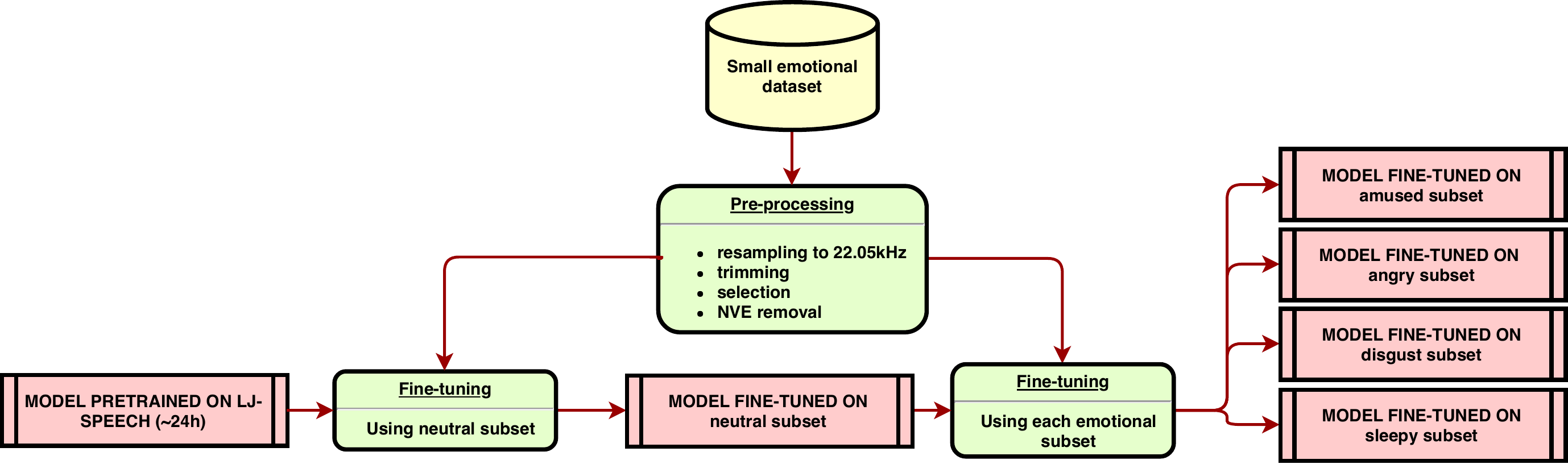}
\caption{Block diagram of the system}
\label{system}
\end{figure*}


Our goal is to study the feasibility of fine-tuning a TTS system pre-trained on a big dataset on few new data and analyze how much the model is able to fit them.
This Section describes the whole system. Figure~\ref{system} represents its overall idea. First, in Section~\ref{tts}, we present the TTS system used as a basis for fine-tuning. We then briefly present the dataset we are using in Section~\ref{dataset}. In Section~\ref{prepro}, we explain the pre-processing of our dataset. Finally, in Section~\ref{finetune}, we detail the fine-tuning procedure applied to obtain emotional TTS models.



\subsection{TTS System}
\label{tts}

The number of deep learning-based TTS system of the state of the art are growing. To carry out our experiments, we chose, DCTTS~\cite{dctts-17-tachibana}, a system that seems to combine advantages of several systems. DCTTS models a sequence-to-sequence problem with a encoder-decoder structure along with an Attention Mechanism like Tacotron~\cite{tacotron-17-wang}. However, unlike Tacotron, the modules of the architecture are all CNN-based and there is no RNN component. In ~\cite{dctts-17-tachibana}, they compared an open source implementation of Tacotron to DCTTS and have higher Mean Opinion Score~(MOS).

In this work, we use the Tensorflow implementation provided in~\cite{dctts-18-kyubyong}.


There are two modules trained separately: Text2Mel and SSRN (for Spectrogram Super-resolution Network).
Text2Mel does the mapping between character embeddings and the output of Mel Filter Banks (MFBs) applied on the audio signal, that is, a mel-spectrogram. Then the second module SSRN does the mapping between the mel-spectrogram and full resolution spectrogram. Finally, Griffin-Lim is used as a vocoder.

Text2Mel module models the sequence-to-sequence task. It is composed of a Text  Encoder,  an Audio  Encoder,  an Attention Mechanism,  and  an Audio  Decoder. 

\subsection{Dataset Used}
\label{dataset}
The dataset used in this work is EmoV-DB: The Database of Emotional Voices~\cite{emov-db-18-tits} that is available online~\footnote{https://github.com/numediart/EmoV-DB}.

EmoV-DB contains sentences uttered by male and female actors in English and a male actor in French. The English actors were recorded in 2 different anechoic chambers of the Northeastern University campus while the French actor was recorded in an anechoic chamber of the University of Mons.

Each actor was asked to utter a subset of the sentences from the CMU-Arctic~\cite{cmuArctic04speechDB} database for English speakers and from the SIWIS~\cite{siwis17speechDB} database for the French speaker. The actors uttered these sentences with 5 emotion classes making it possible to build synthesis and voice transformation systems. For every speaker, the different emotions were recorded in different sessions.

In this work we used one of the English actress to perform emotion adaptation of the TTS system.

The experiments performed on this dataset also assess its usability with deep learning algorithms for voice generation systems.


\subsection{Pre-processing}
\label{prepro}
Important aspects of the pre-processing to use this model are 

\begin{itemize}
\item the sample frequency
\item the trimming of silences at the beginning and end of audio files
\item removal of non-verbal expressions (laughters, yawns, etc.)
\end{itemize}

As the model was trained with LJ-speech database with a sample frequency of 22050Hz, we should use the same with our database. 

The trimming of silences is important because the model use guided attention~\cite{dctts-17-tachibana}. It helps the attention mechanism by assuming that the ordering of characters is almost linearly related to the time in the audio file. This is true only if the speech begins from the start of the file without a silence. 

We experimented that without this trimming, the synthesized sentences often omitted the first words of the text to pronounce. The implementation of~\cite{dctts-18-kyubyong} already use trimming with librosa library. However we noticed that default parameters were not suited for our database and changed $top\_db$ to 20dB.

The same problem happens when there are non verbal expressions in the audio files such as laughters, yawns or sighs. Indeed in these cases, the hypothesis to use guided attention is not verified as well. To overcome this problem, we first manually selected utterances without such non verbal expressions (NVE) for the amused dataset (156 utterances) and the sleepy dataset (361 utterances). Then for the amused dataset only, we augmented this selection by manually removing laughters from a part of the remaining utterances (82 utterances) to have a total of 238 utterances because the selection was quite small.

\subsection{Fine-tuning}
\label{finetune}
In this Section, we explain how we leveraged knowledge transfer on TTS by fine-tuning a part of a pre-trained model on our small dataset.

The pre-training of the model was done using the LJ-Speech dataset. This dataset is available online\footnote{https://keithito.com/LJ-Speech-Dataset} and contains 23.9h of speech uttered by a single female speaker.

The fine-tuning was done with the dataset described in Section \ref{dataset}.

There are several possibilities of how we could fine-tune the model. First, we can choose which parts of the pre-trained model we want to fine-tune with the new dataset and which part we want to keep fixed.

The second part of the model, SSRN, does the mapping between MFBs and full spectrogram. Therefore, it should not depend on the speaker identity on speaking style as it is just trained to do the mapping between two audio features. However, as the model has been pre-trained on one speaker, there is a possibility of over-fitting on the characteristics of that specific speaker. The first question we want to answer is whether the SSRN can generalize the mapping to other speaking styles.

As for Text2Mel module, it is composed of a Text  Encoder,  Audio  Encoder,  Attention,  and  Audio  Decoder. As the text does not depend on characteristics of the speaker or his speaking style, we tried to train only the Audio part. However we found that there are some problems of rhythm in synthesized speech. We believe this is because Attention module is not adapted to the new speaking style. As a consequence, we chose to fine-tune on the entire Text2Mel module.

\section{EXPERIMENT}

In this Section, we detail the experiments performed on the system. 

In the first experiment, we evaluate the usefulness of the fine-tuning technique compared to a random initialization of the parameters of the model. The evaluation is based on a measure of intelligibility of the synthesized speech in terms of word accuracy proposed in~\cite{intellig-16-Orozco}.

In the second experiment, we evaluate the quality of the emotional speech synthesized through a MOS test for each emotion according to the confidence in the perception of the emotion specified.

The amount of speech data used for the experiments are showed in Table \ref{db_durations}. Durations are rounded to the minute. The values between parentheses correspond to the amount of data before selection and NVE removal.

\begin{table}[h]
\caption{Amount of data available for each emotion in terms of total duration and number of utterances}
\label{db_durations}
\begin{center}
\begin{tabular}{|c|c|c|}
\hline
 & Total duration [min] & Number of utterances\\
\hline
Amused & 15 (20) & 238 (296)\\
Angry & 19 & 304\\
Disgusted & 29 & 303\\
Neutral & 23 & 357\\
Sleepy & 36 (51) & 361 (496)\\
\hline
\end{tabular}
\end{center}
\end{table}





\subsection{Objective measures}

\begin{table}[h]
\caption{Intelligibility in terms of Word Accuracy.}
\label{intel}
\begin{center}
\begin{tabular}{|c|c|}
\hline
 & Word Accuracy \\
 \hline
LJ-speech & $0.630 \pm 0.042$ \\ 
\hline
Neutral (random initialization) & $0.004 \pm 0.004$  \\ 
\hline
Neutral (fine-tuning) & $0.517 \pm 0.048$ \\ 

\hline
\end{tabular}
\end{center}
\end{table}

In this experiment, we synthesized 100 sentences of the Harvard sentences~\cite{harvard_sentences-69-Rothauser} with several models. Then an objective measure of the intelligibility of every sentence was computed in terms of word accuracy~\cite{intellig-16-Orozco}. The measure consists of using an ASR to recognize speech and compute a word accuracy by comparing the result to the text label.
The mean word accuracy with 95\% confidence interval for all models are summarized in Table~\ref{intel}. The first line show the word accuracy of the pre-trained model (using LJ-speech). The second line corresponds to the model trained only on the neutral subset. Finally, the third line corresponds to the pre-trained model fine-tuned on the neutral subset.

These measures allow us to compare the fine-tuning of the parameters of a pre-trained model with the random initialization of the parameters of the model. 

The experiments clearly shows that model trained with the neutral subset of 20 min is unable to generate intelligible speech if the parameters are randomly initialized. However, if the initialization of the parameters

\subsection{Perception tests}

After fine-tuning from the neutral model to emotional models, we synthesized 5 sentences not seen during training with each of these models. These sentences were used in a Mean Opinion Score~(MOS) test. 

During this test, the participants were asked to complete a form.


This survey contained 5 sections. Each section was dedicated to one emotion. In every section, the participants were asked to rate utterances between 0 and 5 for the confidence in the perception of the emotion specified (0=we can not hear the emotion specified, 5=we perfectly hear the emotion specified).

This test was performed on both original files from the dataset and synthesized files. Table~\ref{mos_orig} gives MOS with 95\% confidence interval for the original files and Table~\ref{mos_synth} gives them for the synthesized files.

\begin{table}[h]
\caption{MOS test results of original files}
\label{mos_orig}
\begin{center}
\begin{tabular}{|c|c|}
\hline
  & Confidence\\
\hline
Amused & $4.60 \pm 0.20$\\ 
Angry &  $4.22 \pm 0.25$\\ 
Disgusted &  $3.28 \pm 0.27$\\ 
Neutral &  $4.37 \pm 0.23$\\ 
Sleepy &  $3.80 \pm 0.27$\\ 

\hline
\end{tabular}
\end{center}
\end{table}

\begin{table}[h]
\caption{MOS test results of synthesized files}
\label{mos_synth}
\begin{center}
\begin{tabular}{|c|c|}
\hline
  & Confidence\\
\hline
Amused &  $2.00 \pm 0.27$
\\ 
Angry & $2.10 \pm 0.28$
\\ 
Disgusted & $2.27 \pm 0.30$ 
\\ 
Neutral &  $3.59 \pm 0.24$
\\ 
Sleepy & $3.29 \pm 0.26$ 
\\

\hline
\end{tabular}
\end{center}
\end{table}




Results of Table~\ref{mos_orig} should be considered as higher bounds as they represent the opinion about original files of the dataset. Results from Table~\ref{mos_synth} should be compared to these higher bounds. One can see that these higher bounds do not meet the maximum value of 5. 

The confidence of the perception of an emotion is altered by the recording quality, the playing quality and the emotion expressed. Moreover, the test was done on non-native English speakers. In~\cite{non_native-06-Alamsaputra}, they conducted an experiment showing a disproportionate disadvantage for the non-native English speaker when listening to synthesized speech compared to their native English speaker. 

In Table~\ref{mos_synth}, for the confidence of the perception of an emotion, the Neutral category has the higher value. A possible explanation of this is that the pre-trained model used has been trained with a neutral corpus and is therefore closer to the Neutral subset used for fine-tuning. For the other emotional categories, the values are a little degraded but still significantly higher than 0 which shows that the emotion is perceptible in the synthesized speech.

\section{CONCLUSIONS AND FUTURE WORK}

In this paper, we present a technique allowing to synthesize emotional speech using a small emotional speech dataset. This technique is based on the fine-tuning of a deep learning-based TTS model with the neutral subset of the small dataset and then fine-tuning the resulting model with each emotional subset to obtain one model per emotional category.

In the first experiment, we show that training the model with random initialization of the parameters gives completely unintelligible speech synthesis. However using a pre-trained model as initialization and fine-tuning it allows to have intelligible speech.

In the second experiment, we show through perception tests that the speech synthesized is correctly perceived as emotional.

To improve these results, we plan to try a multi-speaker and multi-emotional system to be able to share the knowledge of emotional content of several speakers and several emotions of EmoV-DB. The approach in~\cite{tacotron3-18-ye} seems really interesting for this.



\section*{Acknowledgments}


No\'e Tits is funded through a PhD grant from the Fonds pour la Formation \`a la Recherche dans l'Industrie et l'Agriculture (FRIA), Belgium.

\newpage
\bibliographystyle{IEEEbib}
\bibliography{acl2018}

\end{document}